# Topology Detection in Microgrids with Micro-Synchrophasors

Reza Arghandeh[1]  Martin Gahr[2]  Alexandra von Meier[1]  Guido Cavraro[3]  Monika Ruh[2]  Göran Andersson[2]
Member, IEEE  Student, IEEE  Member, IEEE  Student, IEEE  Member, IEEE  Fellow, IEEE

[1] California Institute for Energy and Environment
University of California-Berkeley
Berkeley, USA

[2] Power Systems Laboratory
ETH Zurich
Zürich, Switzerland

[3] DEI - University of Padova
Padova, Italy

*Abstract*— Network topology in distribution networks is often unknown, because most switches are not equipped with measurement devices and communication links. However, knowledge about the actual topology is critical for safe and reliable grid operation. This paper proposes a voting-based topology detection method based on micro-synchrophasor measurements. The minimal difference between measured and calculated voltage angle or voltage magnitude, respectively, indicates the actual topology. Micro-synchrophasors or micro-Phasor Measurement Units (µPMU) are high-precision devices that can measure voltage angle differences on the order of ten millidegrees. This accuracy is important for distribution networks due to the smaller angle differences as compared to transmission networks. For this paper, a microgrid test bed is implemented in MATLAB with simulated measurements from µPMUs as well as SCADA measurement devices. The results show that topologies can be detected with high accuracy. Additionally, topology detection by voltage angle shows better results than detection by voltage magnitude.

## I. INTRODUCTION

Because of the scale of power flowing through transmission systems, it has been necessary and economical to equip these networks with measurement devices at every node. In practice, the ratio of measurements to system state variables in transmission networks is roughly 1.7 to 2.2 [1]. At the distribution level, the ratio of measurements to state variables is much lower. Distribution networks, historically speaking, are poorly monitored and often managed manually, by sending crews to change switch status on location. To the distribution operator, the real-time status of switching devices may be unknown or uncertain. While reconfiguration actions are more frequent than in transmission networks, communication links to switching devices are limited and manipulations by maintenance teams might go unreported. However, with the integration of distributed energy resources (DER), electric vehicles and controllable loads, reliable knowledge about the topology status of distribution networks becomes more critical for safety, efficiency, and to prevent constraint violations.

There exists a need, therefore, for methods to independently verify the connectivity of the system in real-time.

There exist several different approaches to solve the topology detection problem in distribution networks. The most common approach is based on the minimization of a state estimator residual (error) [2]. Authors of [3] proposed a state estimation algorithm that incorporates switching device status as additional state variables. Both methods are based on the weighted least-squares (WLS) state estimation, which has convergence difficulty in distribution networks due their mostly radial structure. Another class of topology detection algorithms is based on equivalent impedance at feeder levels [4]. The impedance-based approaches cannot guarantee accurate topology detection in distribution networks, however, because different topologies along with variable loads in the radial structure can appear as having similar impedances.

In this paper, a topology detection approach based on multiple synchrophasors or phasor measurement units (PMU) is proposed. The main idea is to take advantage of time-series measurement data. This is a voting-based posterior processing approach inspired by high-precision micro-synchrophasors or µPMUs in whose development the authors are involved [5]. Analyses presented in this paper indicate that voltage phase angle measurements yield more conclusive information than voltage magnitude measurements about the actual grid topology. Another advantage of the proposed algorithm is its ease of implementation. Three voting-based approaches are applied in this paper to all PMUs installed on a hypothetical microgrid test feeder to increase the accuracy of topology detection.

The rest of this paper is organized as follows: Section II and Section III present network and measurement models. Section IV presents the topology detection algorithm. Section V shows simulation results, and Section VI offers conclusions.

## II. DISTRBUTION NETWORK MODEL

This section presents the microgrid used in this paper as the test bed for the case study. The microgrid is modeled as a graph $G=\{V,E\}$, where $V=\{1,2,...,N\}$ are vertices or buses of the network and $\varepsilon=\{1,2,...,E\}$ are edges or lines of the network. In the incidence matrix $A \in \{-1, 0, 1\}^{E \times N}$, the *l*th row represents the line *l* where all elements in that row are zero except the source node (-1) and the terminal node (1). We define the bus admittance matrix **Y** as follow:

$$[Y]_{ij} = \begin{cases} \sum_{k \neq j} y_{kj}, & \text{if } k = j \\ -y_{ij}, & \text{otherwise} \end{cases} \quad k,j \in \varepsilon^{1 \times E} \quad (1)$$

where *i* and *j* are buses of the grid. Kirchhoff's laws at all nodes and lines of the grid satisfy the following equations:

$$\mathbf{I} = \mathbf{Y} \cdot \mathbf{V} \quad (2)$$

$$\mathbf{V_0} = \mathbf{V_N} \quad (3)$$

where **I** and **V** are vectors for current and voltage phasors for all nodes. $\mathbf{V_0}$ is a vector for initial voltage value at all nodes. $\mathbf{V_N}$ is a vector that all its elements are nominal voltage value at substation. We assume the substation has the nominal voltage level $V_N$. We model all nodes as constant power or PQ-buses except the substation, which is modeled as an ideal voltage generator or slack bus. Injected active and reactive power at each node *k* are

$$i_k \cdot v_k = S_k = P_k + jQ_k \quad (4)$$

The power flow equations are developed and solved using the Newton-Raphson method[6].

The case study in this paper is a five-bus microgrid characterized by **B** = *{B1, B2, B3, B4, B5}* with five lines **Line** = *{L12, L13, L24, L34, L35}* equipped with switches **SW** = *{S12, S13, S24, S34, S35}*. In this paper, for the sake of simplicity, each line has one switch. In reality, each line may have to switches at its either end. Five different topologies can are possible in this microgrid. Figure 1 shows a schematic of the physical topology.

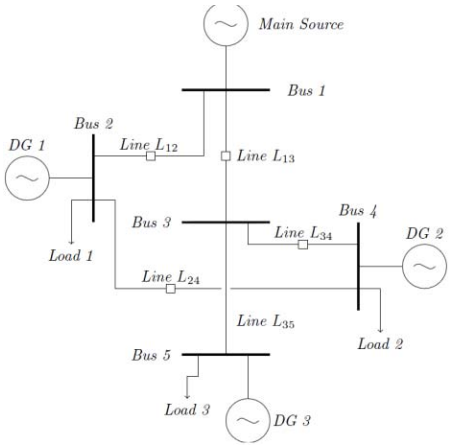

Figure 1. Schematic of the physical topology of the microgrid.

Table I shows all possible topologies considered in the microgrid. Topology V is a meshed network; all other topologies are purely radial. Table II shows the microgrid characteristics. Each bus is equipped with a µPMU. Loads and DGs are monitored with supervisory control and data acquisition (SCADA) measurements.

TABLE I. DIFFERENT TOPOLOGIES FOR THE CASE STUDY

| Topology | Line Status | |
|---|---|---|
| | Connected | Disconnected |
| I | $L_{12}, L_{13}, L_{34}, L_{35}$ | $L_{24}$ |
| II | $L_{13}, L_{24}, L_{34}, L_{35}$ | $L_{12}$ |
| III | $L_{12}, L_{24}, L_{34}, L_{35}$ | $L_{13}$ |
| IV | $L_{12}, L_{13}, L_{24}, L_{35}$ | $L_{34}$ |
| V | $L_{12}, L_{13}, L_{24}, L_{34}, L_{35}$ | – |

TABLE II. MICROGRID BUS CHARACTRISTICS

| Bus Nr. | Typ | Voltage $V$ [p.u.] | Voltage angle $\theta$ [°] | Load $P,Q$ [p.u.] | Generation $P,Q$ [p.u.] |
|---|---|---|---|---|---|
| 1 | Slack Bus | Set to 1p.u. | Set to 0° | No Load | Unknown |
| 2 | Load Bus | Mesaured | Mesaured | Household load | DG (PV) |
| 3 | Load Bus | Mesaured | Mesaured | No Load | No Generation |
| 4 | Load Bus | Mesaured | Mesaured | Household load | DG (PV) |
| 5 | Load Bus | Mesaured | Mesaured | Industrial load | DG (PV) |

Table III shows line impedance for the case study.

TABLE III. LINE IMPEDANCES

| Line | Length L [km] | Impedance [p.u.] |
|---|---|---|
| $L_{12}$ | 8 | 0.009 + j 0.011 |
| $L_{13}$ | 4.5 | 0.005 + j 0.006 |
| $L_{24}$ | 16.625 | 0.019 + j 0.022 |
| $L_{34}$ | 9 | 0.011 + j 0.012 |
| $L_{35}$ | 4.5 | 0.005 + j 0.006 |

## III. MICRO-SYNCHROPHASOR MESAUREMENTS MODEL

The topology detection approach in this paper is inspired by µPMUs developed specifically for distribution networks, where phase angle differences between nodes are two orders of magnitude smaller than in transmission networks, where a resolution down to 0.01° is required [5, 7]. The TVE is smaller than 0.05 %. We assume that the measurement at node *n* and time *t* has the following format:

$$V_n(t) = |v_n(t)| + w_n^V(t), \quad E\left[\left(w_n^V(t)\right)^2\right] = a^2 V_0^2 \quad (5)$$

$$\theta_n(t) = |v_n(t)| + w_n^\theta(t), \quad E\left[\left(w_n^\theta(t)\right)^2\right] = b^2 \quad (6)$$

where *a* and *b* are the standard deviation of voltage magnitude and phase angle measurement due to measurement noise with the µPMU. For the sake of simplicity, we assume $a = b = \sigma_{\mu PMU}$

In addition to µPMU measurements, SCADA measurements are also considered in the case study as legacy measurements.

## IV. TOPOLOGY DETECTION METHOD

The main idea derives from the fact that time-series phasor measurement data from µPMU show specific patterns regarding the current network topology. The proposed method is based on comparing µPMU measurements with calculated system states.

### A. Algorithm Structure

The algorithm is based on minimizing the difference of measurements and calculated values for the system states. The

Newton-Raphson power flow approach is used to calculate the system states for different possible topologies. A library of calculated system states is created as a benchmark for comparison of µPMU measurements. The hypothesis is that the smallest difference between power flow results and measurements infers the correct topology, suggesting that at every bus and every time step, the correct topology can be determined with this approach. The topology detection algorithm has the following steps:

1) µPMUs at the buses $i \in B_{pmu} = \{1, ..., N_{pmu}\} \subseteq B^{1 \times N_{bus}}$ measure voltage magnitude and phase angle.

2) The algorithm builds the measurement matrix $y \in \mathbb{R}^{N_{pmu} \times 2}$, whereby the number of µPMU measurements is $N_{meas} = 2 \times N_{pmu}$.

3) The library of possible topologies is built with running the power flow analysis. The number of possible topologies is $N_T$.

4) The differences between measurements and topology library values are calculated for each µPMU.

$$\Delta\theta_{p,q} = \theta_{PMU,p} - \theta_{PF,q} \quad (7)$$

µPMU measurement: p, p = {1, 2,...,$N_{pmu}$}

Calculated system state for topology: q, q = {1, 2,...,$N_T$}

5) An angle difference matrix **ADM** and a magnitude difference matrix **MDM** are created to include the phase angle difference for each µPMU and its associated values from different possible topologies. Each row represents the difference of specific µPMU measured phase angles with different possible topologies.

$$\mathbf{ADM} = \begin{bmatrix} \Delta\theta_{1,1} & \cdots & \Delta\theta_{1,N_T} \\ \vdots & \ddots & \vdots \\ \Delta\theta_{N_{pmu},1} & \cdots & \Delta\theta_{N_{pmu},N_T} \end{bmatrix}_{N_{pmu} \times N_T} \quad (8)$$

$$\mathbf{MDM} = \begin{bmatrix} \Delta V_{1,1} & \cdots & \Delta V_{1,N_T} \\ \vdots & \ddots & \vdots \\ \Delta V_{N_{pmu},1} & \cdots & \Delta V_{N_{pmu},N_T} \end{bmatrix}_{N_{pmu} \times N_T} \quad (9)$$

6) The **ADM** and **MDM** are the bases for quantifying the similarity of measurements with possible topologies. However, the topology detection accuracy depends on the voting-based topology detection approach. Different approaches for topology detection are explained in the next section.

7) Each row of the **ADM** or **MDM** is associated with one µPMU measurement. After selecting a favored topology for each row of the **ADM**, a voting schema is used to select the highest rated topology.

### B. Voting-based Detection Criteria

In this paper, three different approaches are introduced for topology detection. The second and third approaches are based on voting for all installed µPMU measurements. These approaches are row operators on **ADM** and **MDM** matrices.

*1) Row Minimum Value (RMV):* The current topology (AT) is indicated by the minimum delta in every row of the **ADM** (or **MDM**).

$$AT_i = argmin\{\Delta\theta_{i,1}, ..., \Delta\theta_{i,q}, ..., \Delta\theta_{i,N_T}\} \quad (10)$$

$i \in N_{pmu}$

The RMV can lead to discrepancies, because different rows can have minimum values on different columns. It means different µPMUs may converge to different topologies, in which case the results are inconclusive.

*2) Average Row Minimum Value (ARMV):* The actual topology (AT) is indicated by averaging the minimum delta in each row of the **ADM** (or **MDM**).

$$AT = argmin\{\frac{1}{N_{pmu}}\sum_{q=1}^{N_T}\sum_{i=1}^{N_{pmu}}\Delta\theta_{i,q}\} \quad (11)$$

$i \in N_{pmu}$

*3) Overal Row Minimum Value (ORMV):* Only if all buses indicate the same topology as the correct topology, the result is used. This would mean that the minimum of each row has to be in the same column for all rows in **ADM** (or **MDM**).

$$AT_i = argmin\{\Delta\theta_{i,1}, ..., \Delta\theta_{i,q}, ..., \Delta\theta_{i,N_T}\} \quad (12)$$

$\forall i: AT_1 = ... = AT_{N_{pmu}}$

$i \in N_{pmu}$

All three approaches use the minimum delta as an indicator for the correct topology. However, one could also have chosen a formal threshold; once the delta falls below this threshold it indicates a topology. Yet, taking the minimum seems to be more straightforward. This study is focused on voltage phase angle data and the advantages of µPMU measurements for topology detection. However, the alternative approach can be a combination of voltage magnitude and phase angle for topology detection error are considered as a Total Vector Error (TVE) index.

### V. SIMULATIONS AND DISCUSSIONS

The microgrid simulation setup includes power flow calculation and the creating measurement signals.

### A. Mesaurement Simulation

Measurements from µPMUs and SCADA measurement devices are inputs to the topology detection algorithm. In this paper, synthetic measurements are used to validate the algorithm. The synthetic measurement data are derived from power flow results with added uncertainty and noise. The accuracy is related to the measurement device and is a relative error to the measurand range. It shows the distance of the measurements from true values, proportionally to the measurand magnitude. The measurement noise might originate from the physical sensors, data processing, communication, or some other unknown reasons.

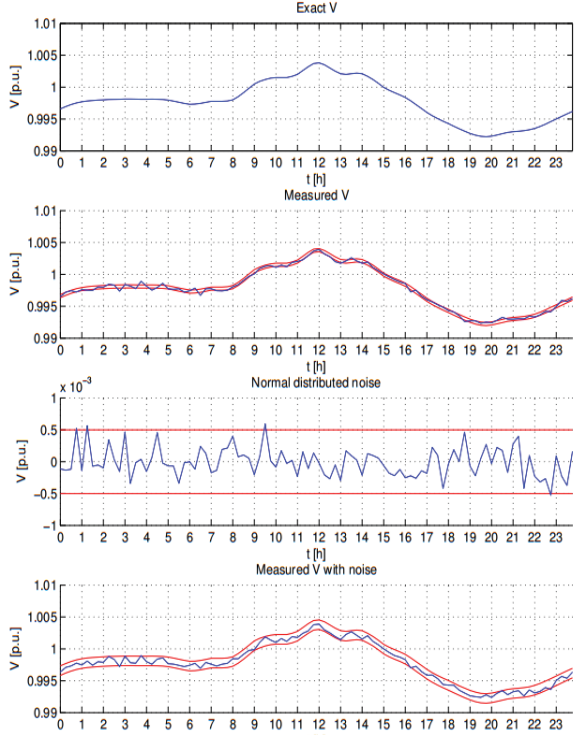

Figure 2. Example of a measurement simulation.

We assume this noise to be independent of the measurands and have a Gaussian distribution $\mathcal{N}(\mu = 0, \sigma)$. We further assume a standard deviation for the µPMU of $\sigma = \pm 0.025\%$, and $\sigma = \pm 2.5\%$ for the SCADA measurements. The measurement accuracy for the µPMU measurements is $\alpha = \pm 0.025\%$ and for the SCADA measurements is $\alpha = \pm 0.05\%$. Figure 2 shows an example of measurement simulation.

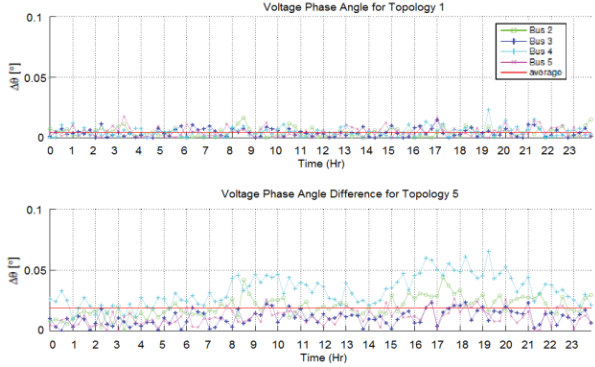

Figure 3. Comparing Δθ values for topology I and V with µPMU measurements.

### B. Microgrid Power Flow Simulation

The case study is a microgrid with five buses which all are equipped with µPMU. The load data are based on two load classes for residential and commercial loads in Switzerland. Both have 15-minute measurements for a weekday in spring. PV generation data are applied for the same location. Power flow analysis and measurement simulation are incorporated into the topology detection algorithm as presented in the previous section. In this study 96 power flow time values are calculated for each topology based on the available load data. However, the algorithm is scalable to various loads and different number of µPMUs in a microgrid based on available load measurements.

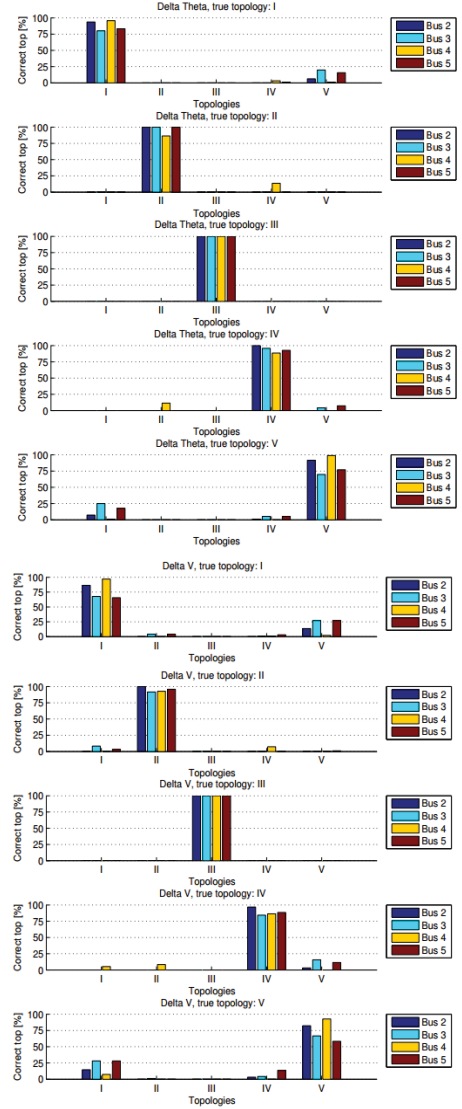

Figure 4. Rate of correct detected topologies with Δθ and ΔV.

As a sample case for validating the algorithm, we assume Topology I is the true topology and examine the algorithm output for the true and for another, wrong topology. The output consists of the values of the angle difference and the magnitude difference matrices in the rows corresponding to each topology. Figure 3 shows these values of the **ADM** and **MDM** rows for Topology I and Topology V. As expected, it shows that the true topology has a lesser difference Δθ and ΔV on average than the wrong topology. This hypothesis is tested for the 96 time values with different loads considering voltage magnitude and phase angle delta values. The exercise is repeated for each possible topology as the true one.

Figure 4 illustrates the topology detection results for all nodes at every time step. The normalized frequency of correctly detected topologies is presented for each bus separately for voltage magnitude and voltage phase angle. It shows that the algorithm detects the true topology with high accuracy. The most challenging condition is distinguishing Topology I from Topology V. Topology V is the meshed topology that is created by closing switch S24. Since the status of switch S24 and thus the resulting power flow on Line L24 are the only differences between Topology I and Topology V, these two topologies are so similar that it is difficult for the algorithm to correctly distinguish between them at all times. However, the algorithm still correctly distinguished between topology I and V in 85% of the cases.

Figure 5 shows the rates of correct detection based on only the simplest index, the RMV, for all topologies. . It shows Bus 5 values for topology I and V are very close based on the RWM approach. To address this ambiguity, we will also examine two other approaches for voting-based topology detection from different μPMUs, which perform better for certain cases.

Figure 5 also shows how topology detection based on voltage phase angle (ADM) is more accurate than voltage magnitude (MDM). Moreover, the success rate of correctly detected topologies depends on the location of μPMUs – specifically, their proximity to a switch whose status is to be ascertained. For example, μPMUs at the frequency of correct detection for Topology V is less for bus 3 and bus 5 μPMUs. Topology V is similar with all other possible topologies expeted one switch state or rather one line not been used.

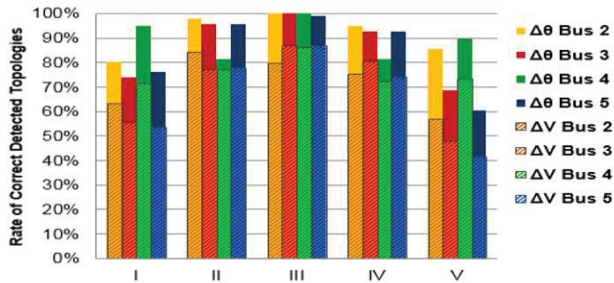

Figure 5. Rate of correctly detected topologies based on RMV criterion

The rate of correctly detected topologies with the ARMV approach is presented in Figure 6. It is based on averaging over all μPMU measurements, and identifying the topology with the minimum average error as the true topology. As can be seen in Figure 6, ARMV performs better than RMW, yielding success rates in the high 90s of percent for voltage angle measurements.

The third approach is ORMV, where the criterion for identifying the correct topology is an agreement by all μPMUs on the same topology. As illustrated in Figure 7, this approach is less successful than the ARMV, since there are many cases in which all μPMUs do not converge to one topology. Therefore, we find the ARMV to be the best among the three detection approaches.

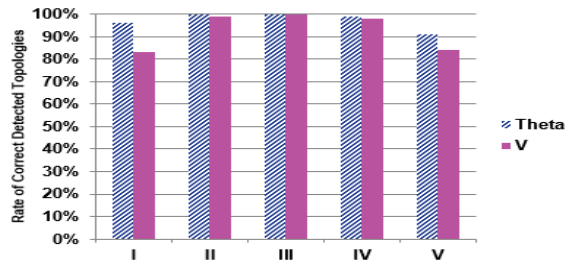

Figure 6. Rate of correctly detected topologies based on ARMV criterion

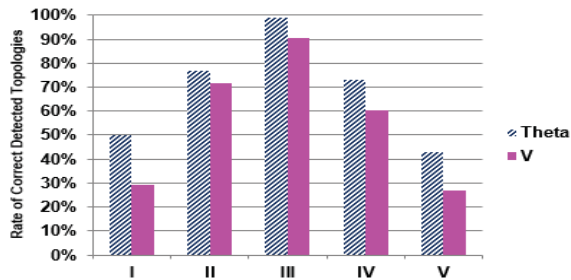

Figure 7. Frequency of correct detect topologies based on ORMV criterion

## VI. COCLUSIONS AND FUTURE WORKS

In this paper, a voting-based topology detection method is presented that uses measurements from high-precision phasor measurement units or μPMUs, an emerging technology to support visibility and situational awareness in distribution networks. The proposed algorithm works with voltage magnitude and voltage angle measurements. The analysis presented here showed that voltage angle measurements led to a more accurate topology detection than voltage magnitude values. Moreover, two different voting-based approaches are examined and compared. The best performing algorithm using the Average Row Minimum Value (ARMV) approach can detect the correct topologies with 85% accuracy in the most challenging cases. The authors are working on enhancing the algorithm accuracy with better μPMU placement and probabilistic voting-based approaches. Assigning TVE thresholds for voting-based approaches will also be addressed in future work.

REFERENCES

[1] A. Monticelli, "Electric power system state estimation," *Proceedings of the IEEE,* vol. 88, pp. 262-282, 2000.
[2] F. F. Wu, *et al.*, "Detection of topology errors by state estimation [power systems]," *Power Systems, IEEE Transactions on,* vol. 4, pp. 176-183, 1989.
[3] G. N. Korres, *et al.*, "A state estimation algorithm for monitoring topology changes in distribution systems," in *Power and Energy Society General Meeting, 2012 IEEE*, 2012, pp. 1-8.
[4] M. Ciobotaru, *et al.*, "On-line grid impedance estimation based on harmonic injection for grid-connected PV inverter," in *Industrial Electronics, 2007. ISIE 2007. IEEE International Symposium on*, 2007, pp. 2437-2442.
[5] A. von Meier, *et al.*, "Micro-synchrophasors for distribution systems," in *Innovative Smart Grid Technologies Conference (ISGT), 2014 IEEE PES*, 2014, pp. 1-5.
[6] J. J. Grainger, *et al.*, *Power system analysis* vol. 621: McGraw-Hill New York, 1994.
[7] A. von Meier, *et al.*, "Diagnostic Applications for Micro-Synchrophasor Measurements," ed, 2014.